\documentstyle[12pt]{article}

\catcode`\@=11
\def\numberbysection{\@addtoreset{equation}{section}
        \def\theequation{\thesection.\arabic{equation}}}
\numberbysection

\input{tcilatex}

\begin{document}

\newlength{\lno} \lno1.5cm \newlength{\len} \len=\textwidth%
\addtolength{\len}{-\lno}

\setcounter{page}{0}

\baselineskip7mm \renewcommand{\thefootnote}{\fnsymbol{footnote}} \newpage %
\setcounter{page}{0}

\begin{titlepage}     
\vspace{0.5cm}
\begin{center}
{\Large\bf A Bethe Ansatz Solution For The Closed $U_{q}[sl(2)]$ Temperley-Lieb Quantum
Spin Chains}\\
\vspace{1cm}
{\large A. Lima-Santos and R.C.T. Ghiotto}\\
\vspace{1cm}
{\large \em Universidade Federal de S\~ao Carlos, Departamento de F\'{\i}sica \\
Caixa Postal 676, CEP 13569-905~~S\~ao Carlos, Brasil}\\
\end{center}
\vspace{1.2cm}

\begin{abstract}
We solve the spectrum of the closed Temperley-Lieb quantum spin chains using
the coordinate Bethe ansatz. These models are invariant under the quantum
group $U_{q}[sl(2)]$.
\end{abstract}

\end{titlepage}

\baselineskip6mm

\newpage{}

\section{Introduction}

Quantum group together with Temperley-Lieb algebra play a important role in
the study of integrable spin chains. It may be interesting to study
particular Hamiltonians associated with the Temperley-Lieb which are
invariant to the quantum group. Taking into account usual toroidal boundary
conditions, the Hamiltonians take the form 
\begin{equation}
{\cal H}=\sum_{n=1}^{N-1}U_{n,n+1}+U_{N1}.  \label{eq1.1}
\end{equation}
where $U_{n,n+1}$ operates in a direct product of two ($2s+1$)-dimensional
complex spaces $V^{2s+1}$ at positions $n$ and $n+1$. They are not invariant
with respect to $U_{q}[sl(2)]$ since $U_{N1}\neq U_{1N}$ breaks
translational invariance, reflecting the non-cocommutativity of the
co-product. Indeed, we know from\cite{Alcaraz, Pasquier, RM} that very
special boundary terms must be considered when we seek these quantum group
invariant spin chains. In particular, one possibility to obtain a quantum
group invariant Hamiltonian is to consider open boundary conditions, {\it %
i.e.}, $U_{N1}=0$ . For The XXZ-Hamiltonian with open boundary conditions
one has to apply the Bethe ansatz techniques introduced by Sklyanin\cite
{Sklyanin} using Cherednik's reflection matrices\cite{Cherednik, MN}. By
this method the XXZ-Heisenberg model\cite{Destri}, the $spl_{q}(2,1)$
invariant supersymmetric t-J model\cite{Angela, Ruiz} , the $U_{q}[sl(n)]$
invariant generalization of the XXZ-chain\cite{Vega} and the $SU_{q}(n|m)$
spin chains \cite{RM, Yue} have been solved for open boundary conditions by
this method.

Recently, by means of a generalized algebraic nested Bethe ansatz, Karowski
and Zapletal\cite{Karowski} presented a class of quantum group invariant $n$%
-state vertex models with periodic boundary conditions. Also an extension of
this method to the case of graded vertex models was analyzed in \cite
{Angela2}, where a $spl_{q}(2|1)$ invariant susy $t$-$J$ model with boundary
conditions was presented.

In fact, this type of models were first discussed by Martin\cite{Martin}
from the representations of the Hecke algebra. The study of closed quantum
group invariant closed spin chains in the framework of the coordinate Bethe
ansatz was presented by Grosse {\it at} {\it al}. for the $SU_{q}(2)$ case 
\cite{Pallua}. In this context it would be interesting to discuss other
quantum group invariant closed spin chains. Therefore, it is the purpose of
this paper to present and solve, via coordinate Bethe ansatz\cite{Bethe} a
closed spin-$s$ Hamiltonian, which in terms of the Temperley-Lieb operators
can be written as

\begin{equation}
{\cal H}=\sum_{n=1}^{N-1}U_{n}+{\cal U}_{0}  \label{eq1.2}
\end{equation}
where 
\begin{equation}
{\cal U}_{0}=GU_{N-1}G^{-1}\quad ,\quad G=(Q-U_{1})(Q-U_{2})\cdots
(Q-U_{N-1})  \label{eq1.3}
\end{equation}
satisfying $[{\cal H},G]=0$ and additionally invariance with respect to the
quantum algebra. The operator $G$ shifts the $U_{n}$ by one unit $%
GU_{n}G^{-1}=U_{n+1}$ and maps ${\cal U}_{0}$ into $U_{1}$, which manifest
the translational invariance of ${\cal H}$.

\section{ The Temperley-Lieb Hamiltonians}

In the basis where $S_{n}^{z}$ is diagonal with eigenvectors $\left|
s,n\right\rangle ,\left| s-1,n\right\rangle ,...,\left| -s,n\right\rangle $
and eigenvalues $s,s-1,...,-s$, the Hamiltonian densities acting on two
neighboring sites are given by 
\begin{equation}
\begin{array}{lll}
\left\langle k,l|\ U\ |i,j\right\rangle & = & \epsilon (i)\epsilon
(k)q^{(i+k)}\delta _{i+j,0}\ \delta _{k+l,0} \\ 
i,j,k,l & = & s,s-1,\cdots ,-s+1,-s
\end{array}
\label{eq1.4}
\end{equation}
where $\epsilon (i)=(-1)^{i}$ for $s$ integer and $\epsilon (i)=(-1)^{i+1/2}$
for $s$ semi-integer. Thus $U_{n}$ denotes the projection on states whose
restriction to sites $n$ and $n+1$ has total spin zero. These Hamiltonians
were derived Batchelor and Kuniba\cite{BK} from representations of the
Temperley-Lieb algebras associated with quantum group $U_{q}[sl(2)]$. The
case $s=1/2$ was investigated in reference\cite{Pallua}.

In fact, $U_{n}$ obeys the Temperley-Lieb algebra\cite{Tlieb} 
\begin{eqnarray}
U_{n}^{2} &=&(Q+Q^{-1})U_{n},\quad Q+Q^{-1}=[2s+1]_{q}  \nonumber \\
U_{n}U_{n+1}U_{n} &=&U_{n},\quad [U_{n},U_{l}]=0\qquad {\rm for\qquad }%
|n-l|\geq 2  \label{eq1.5}
\end{eqnarray}
and commutes with the quantum group $U_{q}[sl(2)]$. The $q$-number notation
is $[x]_{q}=(q^{x}-q^{-x})/(q-q^{-1})$. This algebra appears in a large
class of solvable models and is known to essentially govern their physical
properties: ${\cal H}$ is an element of a set of infinity quantities
conserved which are involutive provided that $U_{n}$ satisfies the defining
relations (\ref{eq1.5}).

Having now built common ground for all closed Hamiltonian densities, whose
salient feature is that they are spin-zero projectors, we may implement the
steps of\cite{RLS} , where the spectrum of the $A$-$D$ Temperley-Lieb
Hamiltonians with either periodic and free boundary conditions were solved ,
via a generalization of the coordinate Bethe ansatz.

\section{The coordinate Bethe ansatz}

Since these Hamiltonians commute with the total spin $S_{T}^{z}=%
\sum_{n=1}^{N}S_{n}^{z}$, the eigenvalues of the operator $r=sN-S_{T}^{z}$
can be used to collect the eigenstates of ${\cal H}$ in sectors, $\Psi _{r}$%
. Due to this $U(1)$ invariance, there always exists a reference state $\Psi
_{0}$ satisfying ${\cal H}\Psi _{0}=E_{0}\Psi _{0}$, with $E_{0}=0$. We take 
$\Psi _{0}$ to be $\Psi _{0}=\prod_{n}\left| s,n\right\rangle $. This is the
only eigenstate in the sector $r=0$. All other energies will be measured
relative to this state.

We will now start to diagonalize ${\cal H}$ in every sector. Nothing
interesting happens in sectors with $r<2s$. Since ${\cal H}$ is a sum of
projectors on spin zero, these states are annihilated by ${\cal H}$.

The first nontrivial sector $r=2s$, the correspondent eigenspace is spanned
by the states $\left| n(-j,j)\right\rangle =\left| s\ s\cdots s\ \stackunder{%
n}{-j}j\ s\cdots s\right\rangle $ , where $n=1,2,...,N-1\ $and{\rm \ }$\
j=-s,-s+1,...,s.$ We seek eigenstates of ${\cal H}$ which are linear
combinations of these vectors. It is very convenient to consider the linear
combination 
\begin{equation}
\left| \Omega (n)\right\rangle =\sum_{j=-s}^{s}(-1)^{s+j}\ q^{s-j}\left|
n(-j,j)\right\rangle .  \label{eq1.6}
\end{equation}
which is a highest weight state, $S^{+}\left| \Omega (n)\right\rangle =0$,
and eigenstate of $U_{n}$%
\begin{eqnarray}
U_{n}\left| \Omega (n)\right\rangle &=&(Q+Q^{-1})\left| \Omega
(n)\right\rangle ,\quad U_{n\pm 1}\left| \Omega (n)\right\rangle =\epsilon
_{s}\left| \Omega (n\pm 1)\right\rangle ,  \nonumber \\
U_{n}\left| \Omega (n\pm 1)\right\rangle &=&\epsilon _{s}\left| \Omega
(n)\right\rangle ,\quad U_{n}\left| \Omega (m)\right\rangle =0\quad {\rm for}%
\quad \quad n\neq \{m\pm 1,m\}  \label{eq1.7}
\end{eqnarray}
where $\epsilon _{s}=-1$ for $s$ semi-integer and $\epsilon _{s}=1$ for $s$
integer. In this basis, all spin-$s$ Hamiltonians ${\cal H}$ can be treated
in a similar way and it affords a considerable simplification in the
diagonalization of ${\cal H}$, when one compares with the computations in
the usual spin basis\cite{RLS}.

\subsection{One-pseudoparticle eigenstates}

Let us consider one free pseudoparticle as a highest weight state which lies
in the sector $r=2s$

\begin{equation}
\Psi _{2s}=\sum_{n=1}^{N-1}A(n)\left| \Omega (n)\right\rangle .
\label{eq1.8}
\end{equation}
Using the eigenvalue equation ${\cal H}\Psi _{2s}=E_{2s}\Psi _{2s}$, one can
derive a complete set of equations for the wavefunctions $A(n)$.

The action of the operator $G=(Q-U_{1})\cdots (Q-U_{N-1})$ on the states $%
\left| \Omega (n)\right\rangle $ can be computed using (\ref{eq1.7}). It is
simple on the bulk and at the left boundary

\begin{equation}
G\left| \Omega (n)\right\rangle =-\epsilon _{s}Q^{N-2}\ \left| \Omega
(n+1)\right\rangle {\rm \quad },{\rm \quad }1\leq n\leq N-2  \label{eq1.9}
\end{equation}
but manifests its nonlocality at the right boundary

\begin{equation}
G\left| \Omega (N-1)\right\rangle =\epsilon
_{s}Q^{N-2}\sum_{n=1}^{N-1}(-\epsilon _{s}Q)^{-n}\ \left| \Omega
(N-n)\right\rangle  \label{eq1.10}
\end{equation}
Similarly, acting with the operator $G^{-1}=(Q^{-1}-U_{N-1})\cdots
(Q^{-1}-U_{1})$ , we get

\begin{equation}
G^{-1}\left| \Omega (n)\right\rangle =-\epsilon _{s}Q^{-N+2}\left| \Omega
(n-1)\right\rangle \ \quad ,\quad 2\leq n\leq N-1  \label{eq1.11}
\end{equation}
\begin{equation}
G^{-1}\left| \Omega (1)\right\rangle =\epsilon
_{s}Q^{-N+2}\sum_{n=1}^{N-1}(-\epsilon _{s}Q)^{n}\ \left| \Omega
(n)\right\rangle  \label{eq1.12}
\end{equation}
for the bulk including the right boundary and for the left boundary,
respectively.

From these results one can see that the action of ${\cal U}%
_{0}=GU_{N-1}G^{-1}$ vanishes on the bulk 
\begin{equation}
{\cal U}_{0}\left| \Omega (n)\right\rangle =0\quad ,\quad 2\leq n\leq N-2
\label{eq1.13}
\end{equation}
and is nonlocal at the boundaries 
\begin{equation}
{\cal U}_{0}\left| \Omega (1)\right\rangle =-\epsilon _{s}\sum_{n=1}^{N-1}\
(-\epsilon _{s}Q)^{n}\ \left| \Omega (n)\right\rangle ,\quad {\cal U}%
_{0}\left| \Omega (N-1)\right\rangle =(-\epsilon _{s}Q)^{-N}\ {\cal U}%
_{0}\left| \Omega (1)\right\rangle  \label{eq1.14}
\end{equation}

Next, the action of the operator ${\cal U}=\sum_{k=1}^{N-1}U_{k}$ on the
states $\left| \Omega (n)\right\rangle $ gives the following equations

\begin{eqnarray}
{\cal U}\left| \Omega (1)\right\rangle &=&(Q+Q^{-1})\left| \Omega
(1)\right\rangle +\epsilon _{s}\left| \Omega (2)\right\rangle  \nonumber \\
{\cal U}\left| \Omega (n)\right\rangle &=&(Q+Q^{-1})\left| \Omega
(n)\right\rangle +\epsilon _{s}\left| \Omega (n-1)\right\rangle +\epsilon
_{s}\left| \Omega (n+1)\right\rangle  \nonumber \\
\qquad \quad {{\rm for }\ }2 &\leq &n\leq N-2  \nonumber \\
{\cal U}\left| \Omega (N-1)\right\rangle &=&(Q+Q^{-1})\left| \Omega
(N-1)\right\rangle +\epsilon _{s}\left| \Omega (N-2)\right\rangle .
\label{eq1.16}
\end{eqnarray}

Before we substitute these results into the eigenvalue equation, we will
define two new states

\begin{equation}
\epsilon _{s}\left| \Omega (0)\right\rangle ={\cal U}_{0}\left| \Omega
(1)\right\rangle ,\quad \ \epsilon _{s}\left| \Omega (N)\right\rangle ={\cal %
U}_{0}\left| \Omega (N-1)\right\rangle  \label{eq1.17}
\end{equation}
to include the cases $n=0$ and $n=N$ into the definition of $\Psi _{2s}$,
equation (\ref{eq1.8}). Finally, the action of ${\cal H}={\cal U}+{\cal U}%
_{0}$ on the states $\left| \Omega (n)\right\rangle $ is

\begin{eqnarray}
{\cal H}\left| \Omega (0)\right\rangle &=&(Q+Q^{-1})\left| \Omega
(0)\right\rangle +(-\epsilon _{s}Q)^{N}\epsilon _{s}\left| \Omega
(N-1)\right\rangle +\epsilon _{s}\left| \Omega (1)\right\rangle  \nonumber \\
&&  \nonumber \\
{\cal H}\left| \Omega (n)\right\rangle &=&(Q+Q^{-1})\left| \Omega
(n)\right\rangle +\epsilon _{s}\left| \Omega (n-1)\right\rangle +\epsilon
_{s}\left| \Omega (n+1)\right\rangle  \nonumber \\
\qquad \qquad {\rm for \ }1 &\leq &n\leq N-2  \nonumber \\
&&  \nonumber \\
{\cal H}\left| \Omega (N-1)\right\rangle &=&(Q+Q^{-1})\left| \Omega
(N-1)\right\rangle +\epsilon _{s}\left| \Omega (N-2)\right\rangle  \nonumber
\\
&&+(-\epsilon _{s}Q)^{-N}\epsilon _{s}\left| \Omega (0)\right\rangle 
\nonumber \\
&&  \nonumber \\
{\cal H}\left| \Omega (N)\right\rangle &=&(Q+Q^{-1})\left| \Omega
(N)\right\rangle +\epsilon _{s}\left| \Omega (N-1)\right\rangle  \nonumber \\
&&+(-\epsilon _{s}Q)^{-N}\epsilon _{s}\left| \Omega (1)\right\rangle
\label{eq1.18}
\end{eqnarray}
Substituting these results into the eigenvalue equation\ ${\cal H}\Psi
_{2s}=E_{2s}\ \Psi _{2s}$ and using the boundary conditions 
\begin{equation}
(-\epsilon _{s}Q)^{N}A(x)=A(N+x){\ }  \label{eq1.19}
\end{equation}
we get a complete set of eigenvalue equations for the wavefunctions

\begin{eqnarray}
E_{2s}\ A(n) &=&(Q+Q^{-1})A(n)+\epsilon _{s}A(n-1)+\epsilon _{s}A(n+1) 
\nonumber \\
\quad \qquad {\rm for\quad }1 &\leq &n\leq N-1  \label{eq1.20}
\end{eqnarray}

The plane wave parametrization $A(n)=\xi ^{n}$ solves these eigenvalue
equations and the boundary conditions (\ref{eq1.19}), provided that: 
\begin{equation}
E_{2s}=Q+Q^{-1}+\epsilon _{s}(\xi +\xi ^{-1})\quad {and}\quad \xi
^{N}=(-\epsilon _{s}Q)^{N}  \label{eq1.21}
\end{equation}
where $\xi ={\rm e}^{i\theta }$ and $\theta $ being the momentum.

\subsection{Two-pseudoparticle eigenstates}

Let us now consider the sector $r=2(2s)$, where we have two interacting
pseudoparticles. We seek the corresponding eigenfunction as products of
single pseudoparticles eigenfunctions, {\it i.e}. 
\begin{equation}
\Psi _{4s}=\sum_{x_{1}+1<x_{2}}A(x_{1},x_{2})\left| \Omega
(x_{1},x_{2})\right\rangle  \label{eq1.22}
\end{equation}
where 
\begin{equation}
\left| \Omega (x_{1},x_{2})\right\rangle
=\sum_{i=-s}^{s}\sum_{j=-s}^{s}(-1)^{i+j}q^{2s-i-j}\left|
x_{1}(-i,i),x_{2}(-j,j)\right\rangle  \label{eq1.23}
\end{equation}

To solve the eigenvalue equation ${\cal H}\Psi _{4s}=E_{4s}\Psi _{4s}$, we
recall (\ref{eq1.7}) to get the action of ${\cal U}$ and ${\cal U}_{0}$ on
the states $\left| \Omega (x_{1},x_{2})\right\rangle $. We have to consider
four cases: (i)\ When the two pseudoparticles are separated in the bulk, the
action of ${\cal U}$ is 
\begin{eqnarray}
{\cal U}\left| \Omega (x_{1},x_{2})\right\rangle &=&2(Q+Q^{-1})\left| \Omega
(x_{1},x_{2})\right\rangle +\epsilon _{s}\left| \Omega
(x_{1}-1,x_{2})\right\rangle +\epsilon _{s}\left| \Omega
(x_{1}+1,x_{2})\right\rangle  \nonumber \\
&&+\epsilon _{s}\left| \Omega (x_{1},x_{2}-1)\right\rangle +\epsilon
_{s}\left| \Omega (x_{1},x_{2}+1)\right\rangle  \label{eq1.24}
\end{eqnarray}
i.e., for $x_{1}$ $\geq 2$ and $x_{1}+3\leq x_{2}\leq N-2$; (ii) When the
two pseudoparticles are separated but one of them or both are at the
boundaries 
\begin{eqnarray}
{\cal U}\left| \Omega (1,x_{2})\right\rangle &=&2(Q+Q^{-1})\left| \Omega
(1,x_{2})\right\rangle +\epsilon _{s}\left| \Omega (2,x_{2})\right\rangle
+\epsilon _{s}\left| \Omega (1,x_{2}-1)\right\rangle  \nonumber \\
&&+\epsilon _{s}\left| \Omega (1,x_{2}+1)\right\rangle  \label{eq1.25}
\end{eqnarray}
\begin{eqnarray}
{\cal U}\left| \Omega (x_{1},N-1)\right\rangle &=&2(Q+Q^{-1})\left| \Omega
(x_{1},N-1)\right\rangle +\epsilon _{s}\left| \Omega
(x_{1}-1,N-1)\right\rangle  \nonumber \\
&&+\epsilon _{s}\left| \Omega (x_{1}+1,N-1)\right\rangle +\epsilon
_{s}\left| \Omega (x_{1},N-2)\right\rangle  \label{eq1.26}
\end{eqnarray}
\begin{equation}
{\cal U}\left| \Omega (1,N-1)\right\rangle =2(Q+Q^{-1})\left| \Omega
(1,N-1)\right\rangle +\epsilon _{s}\left| \Omega (2,N-1)\right\rangle
+\epsilon _{s}\left| \Omega (1,N-2)\right\rangle  \label{eq1.27}
\end{equation}
where $2\leq x_{1}\leq N-4$ and $4\leq x_{2}\leq N-2$; (iii) When the two
pseudoparticles are neighbors in the bulk 
\begin{eqnarray}
{\cal U}\left| \Omega (x,x+2)\right\rangle &=&2(Q+Q^{-1})\left| \Omega
(x,x+2)\right\rangle +\epsilon _{s}\left| \Omega (x-1,x+2)\right\rangle
+\epsilon _{s}\left| \Omega (x,x+3)\right\rangle  \nonumber \\
&&+U_{x+1}\left| \Omega (x,x+2)\right\rangle  \label{eq1.28}
\end{eqnarray}
for $2\leq x\leq N-4$ and (iv) When the two pseudoparticles are neighbors
and at the boundaries 
\begin{equation}
{\cal U}\left| \Omega (1,3)\right\rangle =2(Q+Q^{-1})\left| \Omega
(1,3)\right\rangle +\epsilon _{s}\left| \Omega (1,4)\right\rangle
+U_{2}\left| \Omega (1,3)\right\rangle  \label{eq1.29}
\end{equation}
\begin{eqnarray}
{\cal U}\left| \Omega (N-3,N-1)\right\rangle &=&2(Q+Q^{-1})\left| \Omega
(N-3,N-1)\right\rangle +\epsilon _{s}\left| \Omega (N-4,N-1)\right\rangle 
\nonumber \\
&&+U_{N-2}\left| \Omega (N-3,N-1)\right\rangle  \label{eq1.30}
\end{eqnarray}

Moreover, the action of ${\cal U}_{0}$ does not depend on the
pseudoparticles are neither separated nor neighbors. It is vanishes in the
bulk 
\begin{equation}
{\cal U}_{0}\left| \Omega (x_{1},x_{2})\right\rangle =0\quad {{\rm for}\quad 
}x_{1}\neq 1\ {\rm and\quad }x_{2}\neq N-1,  \label{eq1.31}
\end{equation}
and different of zero at the boundaries: 
\begin{eqnarray}
{\cal U}_{0}\left| \Omega (1,x_{2})\right\rangle  &=&-\epsilon
_{s}\sum_{k=1}^{x_{2}-2}(-\epsilon _{s}Q)^{k}\left| \Omega
(k,x_{2})\right\rangle -(-\epsilon _{s}Q)^{x_{2}-1}U_{x_{2}}\left| \Omega
(x_{2}-1,x_{2}+1)\right\rangle   \nonumber \\
&&-\epsilon _{s}\sum_{k=x_{2}+2}^{N-1}(-\epsilon _{s}Q)^{k-2}\left| \Omega
(x_{2},k)\right\rangle   \label{eq1.32}
\end{eqnarray}
\begin{equation}
{\cal U}_{0}\left| \Omega (x_{1},N-1)\right\rangle =(-\epsilon
_{s}Q)^{-N+2}\ {\cal U}_{0}\left| \Omega (1,x_{2})\right\rangle 
\label{eq1.33}
\end{equation}
where $2\leq x_{1}\leq N-3$ and $3\leq x_{2}\leq N-2$.

Before we substitute these expressions into the eigenvalue equation, we
define new states in order to have consistency between bulk and boundaries
terms 
\begin{eqnarray}
{\cal U}_{0}\left| \Omega (1,x_{2})\right\rangle &=&\epsilon _{s}\left|
\Omega (0,x_{2})\right\rangle ,\quad {\cal U}_{0}\left| \Omega
(x_{1},N-1)\right\rangle =\epsilon _{s}\left| \Omega (x_{1},N)\right\rangle 
\nonumber \\
{\cal U}_{0}\left| \Omega (1,N-1)\right\rangle &=&\epsilon _{s}\left| \Omega
(0,N-1)\right\rangle +\epsilon _{s}\left| \Omega (1,N)\right\rangle 
\nonumber \\
U_{x+1}\left| \Omega (x,x+2)\right\rangle &=&\epsilon _{s}\left| \Omega
(x,x+1)\right\rangle +\epsilon _{s}\left| \Omega (x+1,x+2)\right\rangle
\label{eq1.34}
\end{eqnarray}
Acting with ${\cal H}$ on these new states, we get 
\begin{eqnarray}
{\cal H}\left| \Omega (0,x_{2})\right\rangle &=&2(Q+Q^{-1})\left| \Omega
(0,x_{2})\right\rangle +\epsilon _{s}\left| \Omega (0,x_{2}-1)\right\rangle
+\epsilon _{s}\left| \Omega (0,x_{2}+1)\right\rangle  \nonumber \\
&&+\epsilon _{s}\left| \Omega (1,x_{2})\right\rangle +(-\epsilon
_{s}Q)^{N-2}\epsilon _{s}\left| \Omega (x_{2},N-1)\right\rangle
\label{eq1.35}
\end{eqnarray}
\begin{eqnarray}
{\cal H}\left| \Omega (x_{1},N)\right\rangle &=&2(Q+Q^{-1})\left| \Omega
(x_{1},N)\right\rangle +\epsilon _{s}\left| \Omega (x_{1}-1,N)\right\rangle
+\epsilon _{s}\left| \Omega (x_{1}+1,N)\right\rangle  \nonumber \\
&&+\epsilon _{s}\left| \Omega (x_{1},N-1)\right\rangle +(-\epsilon
_{s}Q)^{-N+2}\epsilon _{s}\left| \Omega (1,x_{1})\right\rangle
\label{eq1.36}
\end{eqnarray}
\begin{equation}
{\cal H}\left| \Omega (x,x+1\right\rangle =(Q+Q^{-1})\left| \Omega
(x,x+1\right\rangle +\epsilon _{s}\left| \Omega (x-1,x+1\right\rangle
+\epsilon _{s}\left| \Omega (x,x+2\right\rangle  \label{eq1.37}
\end{equation}
Substituting these results into the eigenvalue equation, we get the
following equations for wavefunctions corresponding to the separated
pseudoparticles. 
\begin{eqnarray}
E_{4s}A(x_{1},x_{2}) &=&2(Q+Q^{-1})A(x_{1},x_{2})+\epsilon
_{s}A(x_{1}-1,x_{2})+\epsilon _{s}A(x_{1}+1,x_{2})  \nonumber \\
&&+\epsilon _{s}A(x_{1},x_{2}-1)+\epsilon _{s}A(x_{1},x_{2}+1)
\label{eq1.38}
\end{eqnarray}
{\it i.e}., for $x_{1}\geq 1$ and $x_{1}+3\leq x_{2}\leq N-1$. The boundary
conditions read now 
\begin{equation}
A(x_{2},N+x_{1})=(-\epsilon _{s}Q)^{N-2}A(x_{1},x_{2}).  \label{eq1.39}
\end{equation}
The parametrization for the wavefunctions

\begin{equation}
A(x_{1},x_{2})=A_{12}\xi _{1}^{x_{1}}\xi _{2}^{x_{2}}+A_{21}\xi
_{1}^{x_{2}}\xi _{2}^{x_{1}}  \label{eq1.40}
\end{equation}
solves the equation (\ref{eq1.38}) provided that 
\begin{equation}
E_{4s}=2(Q+Q^{-1})+\epsilon _{s}(\xi _{1}+\xi _{1}^{-1}+\xi _{2}+\xi
_{2}^{-1})  \label{eq1.41}
\end{equation}
and the boundary conditions (\ref{eq1.39}) provided that 
\begin{equation}
\xi _{2}^{N}=(-\epsilon _{s}Q)^{N-2}\frac{A_{21}}{A_{12}}\quad ,\quad \xi
_{1}^{N}=(-\epsilon _{s}Q)^{N-2}\frac{A_{12}}{A_{21}}\Rightarrow \xi
^{N}=(-\epsilon _{s}Q)^{2(N-2)}  \label{eq1.42}
\end{equation}
where $\xi =\xi _{1}\xi _{2}=e^{i(\theta _{1}+\theta _{2})}$, $\theta
_{1}+\theta _{2}$ being the total momenta.

Now we include the new states (\ref{eq1.34}) into the definition of $\Psi
_{4s}$ in order to extend (\ref{eq1.22}) to 
\begin{equation}
\Psi _{4s}=\sum_{x_{1}<x_{2}}A(x_{1},x_{2})\left| \Omega
(x_{1},x_{2}\right\rangle .  \label{eq1.43}
\end{equation}
Here we have used the same notation for separated and neighboring states.

Substituting (\ref{eq1.28}) and (\ref{eq1.37}) into the eigenvalue equation,
we get 
\begin{equation}
E_{4s}A(x,x+1)=(Q+Q^{-1})A(x,x+1)+\epsilon _{s}A(x-1,x+1)+\epsilon
_{s}A(x,x+2)  \label{eq1.44}
\end{equation}
which gives us the phase shift produced by the interchange of the two
interacting pseudoparticles 
\begin{equation}
\frac{A_{21}}{A_{12}}=-\frac{1+\xi +\epsilon _{s}(Q+Q^{-1})\xi _{2}}{1+\xi
+\epsilon _{s}(Q+Q^{-1})\xi _{1}}.  \label{eq1.45}
\end{equation}
We thus arrive to the Bethe ansatz equations which fix the values of $\xi
_{1}$ and $\xi _{2}$:

\begin{eqnarray}
\xi _{2}^{N} &=&(-\epsilon _{s}Q)^{N-2}\left\{ -\frac{1+\xi +\epsilon
_{s}(Q+Q^{-1})\xi _{2}}{1+\xi +\epsilon _{s}(Q+Q^{-1})\xi _{1}}\right\} , 
\nonumber \\
\quad \xi _{1}^{N}\xi _{2}^{N} &=&(-\epsilon _{s}Q)^{2(N-2)}  \label{eq1.46}
\end{eqnarray}

\subsection{General eigenstates}

The generalization to any $r$ multiple of $2s$ is in principle
straightforward. Since the Yang-Baxter equations are satisfied, there is
only two-pseudoparticle scattering (if we use the S-matrix language).
Therefore neighbor equations, where more the two pseudoparticles become
neighbors, are not expected to give any new restrictions. For instance, in
the sector $r=3(2s)$ we have three interacting pseudoparticles with
parameters $\xi _{1},\xi _{2}$ and $\xi _{3}$. The corresponding
wavefunctions 
\begin{eqnarray}
A(x_{1},x_{2},x_{3}) &=&A_{123}\xi _{1}^{x_{1}}\xi _{2}^{x_{2}}\xi
_{3}^{x_{3}}+A_{132}\xi _{1}^{x_{1}}\xi _{2}^{x_{3}}\xi
_{3}^{x_{2}}+A_{213}\xi _{1}^{x_{2}}\xi _{2}^{x1}\xi _{3}^{x_{3}}+A_{231}\xi
_{1}^{x_{2}}\xi _{2}^{x_{3}}\xi _{3}^{x_{1}}  \nonumber \\
&&+A_{312}\xi _{1}^{x_{3}}\xi _{2}^{x_{1}}\xi _{3}^{x_{2}}+A_{321}\xi
_{1}^{x_{3}}\xi _{2}^{x_{2}}\xi _{3}^{x_{1}}  \label{eq1.47}
\end{eqnarray}
satisfy the boundary conditions 
\[
A(x_{2},x_{3},N+x_{1})=(-\epsilon _{s}Q)^{N-4}A(x_{1},x_{2},x_{3}) 
\]
which imply that

\begin{equation}
\xi _{i}^{N}=(-\epsilon _{s}Q)^{N-4}\frac{A_{ijk}}{A_{jki}}=(-\epsilon
_{s}Q)^{N-4}\frac{A_{ikj}}{A_{kji}},\quad i\neq j\neq k=1,2,3  \label{eq1.48}
\end{equation}
These relations show us that the interchange of two pseudoparticles is
independent of the position of the third particle. Thus in the sector $%
r=p(2s)$, we expect that the $p$-pseudoparticle phase shift will be a sum of 
$\binom{p}{2}$ two-pseudoparticle phase shifts and the energy is given by 
\begin{equation}
E_{p(2s)}=\sum_{n=1}^{p}\left\{ Q+Q^{-1}+\epsilon _{s}(\xi _{n}+\xi
_{n}^{-1})\right\}  \label{eq1.49}
\end{equation}
where 
\[
\xi _{a}^{N}=(-\epsilon _{s}Q)^{N-2p+2}\prod_{b\neq a}^{p}\left\{ -\frac{%
1+\xi _{a}\xi _{b}+\epsilon _{s}(Q+Q^{-1})\xi _{a}}{1+\xi _{a}\xi
_{b}+\epsilon _{s}(Q+Q^{-1})\xi _{b}}\right\} ,\quad a=1,...,p 
\]
\begin{equation}
\left( \xi _{1}\xi _{2}\cdots \xi _{p}\right) ^{N}=(-\epsilon
_{s}Q)^{p(N-2p+2)}  \label{eq1.50}
\end{equation}

It is not all, in a sector $r$ we may have $p$ pseudoparticle and $%
N_{s-1},N_{s-2},...,N_{-s+1}$ impurities of the type $(s-1),(s-2),...,(-s+1)$%
, respectively, such that 
\begin{equation}
N_{s-1}+2N_{s-2}+\cdots +(2s-1)N_{-s+1}=r-2sp  \label{eq1.51}
\end{equation}
We call impurity a state $\left| a,n\right\rangle $ flanked by at least two
states $\left| b,n\pm 1\right\rangle $ such that $a+b\neq 0$. Since ${\cal H}
$ is a sum of projectors on spin zero, these states are annihilated by $%
{\cal H}$ . In particular, the do not move under the action of ${\cal H}$,
which is the reason for their name. Nevertheless, a pseudoparticle can
propagate past the isolated impurity, but in so doing causes a shift in its
position by two lattice sites. Thus, for a sector $r$ with $l$ impurities
with parameters $\xi _{1},...,\xi _{l}$ and $p$ pseudoparticles with
parameters $\xi _{l+1},...,\xi _{l+p}$ the energy is given by (\ref{eq1.49}%
), and the Bethe equations do not depend on impurity type and are given by 
\begin{equation}
\xi _{a}^{N}\xi _{1}^{2}\xi _{2}^{2}\cdots \xi _{l}^{2}=(-\epsilon
_{s}Q)^{N-2p+2}\prod\Sb b=l+1  \\ b\neq a  \endSb ^{l+p}\left\{ -\frac{1+\xi
_{a}\xi _{b}+\epsilon _{s}(Q+Q^{-1})\xi _{a}}{1+\xi _{a}\xi _{b}+\epsilon
_{s}(Q+Q^{-1})\xi _{b}}\right\}  \label{eq1.52}
\end{equation}
with $a=l+1,l+2,...,l+p\quad ,\quad p\geq 1$, and 
\begin{equation}
\xi ^{2p}(\xi _{l+1}\cdots \xi _{l+p})^{N-2p}=(-\epsilon _{s}Q)^{p(N-2p+2)}
\label{eq1.53}
\end{equation}
where $\xi =\xi _{1}\xi _{2}\cdots \xi _{l}\xi _{l+1}\cdots \xi _{l+p}$.

\section{Conclusion}

We have shown that these closed Temperley-Lieb quantum invariant spin chains
can be solved by the coordinate Bethe ansatz. A consequence of the nonlocal
terms ${\cal U}_{0}$ is the arising of boundary conditions depending on the
quantum group parameter $q$ via the relation $Q+Q^{-1}=[2s+1]_{q}$ and on
the number $p$ of pseudoparticles (which is equal to spin sector $r$ , when $%
s=1/2$).

An interesting extension of this work would be the application of the
methods here presented to solve new strongly correlated electronic systems
associated with the Temperley-Lieb algebras\cite{Links, Angela3}. This is
presently under investigation

{\bf Acknowledgments}: It is ALS pleasure to thank Angela Foerster and
Roland K\"{o}berle for interesting discussions, and CNPq, Brasil, for a
fellowship, providing partial support.

\end{document}